\documentclass[useAMS]{gGAF2e}

\begin{document}
\doi{10.1080/03091920xxxxxxxxx}
 \issn{1029-0419} \issnp{0309-1929} 

\markboth{Nonlinear dynamical modeling of solar cycles}{Nonlinear
dynamical modeling of solar cycles}

\title{Nonlinear Dynamical Modeling of Solar Cycles\\ Using Dynamo
Formulation with Turbulent Magnetic Helicity}
\author{I.N. Kitiashvili$^\ast$$^1$ \& A.G. Kosovichev$^2$\thanks{$^\ast$Corresponding
author. Email: irinasun@stanford.edu\vspace{6pt}
\newline\centerline{\tiny{{\em Geophysical and Astrophysical Fluid
Dynamics}}}
\newline\centerline{\tiny{  ISSN 0309-1929 print/ ISSN 1029-0419 online \textcopyright 2006 Taylor \& Francis Ltd}}
\newline\centerline{\tiny{ http://www.tandf.co.uk/journals}}\newline \centerline{\tiny{DOI:
10.1080/03091920xxxxxxxxx}}}\\\vspace{6pt} $^1$ Center for
Turbulence Research, Stanford University, Stanford, CA 94305, USA\\
$^2$W.W. Hansen Experimental Physics Laboratory, Stanford
University, Stanford, CA 94305, USA}  \received{Received 24 April 2008; in final form 25 July 2008}

\maketitle

\begin{abstract}
Variations of the sunspot number are important indicators of the
solar activity cycles. The sunspot formation is a result of a dynamo
process inside the Sun, which is far from being understood. We use
simple dynamical models of the dynamo process to simulate the
magnetic field evolution and investigate general properties of the
sunspot number variations during the solar cycles. We have found
that the classical Parker's model with a standard kinetic helicity
quenching cannot represent the typical profiles of the solar-cycle
variations of the sunspot number,  and also does not give chaotic
solutions. For modeling of the solar cycle properties we use a
nonlinear dynamo model of Kleeorin and Ruzmaikin (1982), which takes
into account dynamics of the turbulent magnetic helicity. We have
obtained a series of periodic and chaotic solutions for different
layers of the convective zone. The solutions qualitatively reproduce
some basic observational features of the solar cycle properties,
in particular, the relationship between the growth time and the cycle
amplitude. Also, on the longer time scale the dynamo model with the
magnetic helicity has intermittent solutions, which may be important
for modeling long-term variations of the solar cycles.
\end{abstract}

\section{Introduction}

Investigation of solar activity has a long history. The 11-year
sunspot cycles were discovered by Schwabe (1844). Detailed
observations of sunspots locations on the solar disk and their sizes
provided the latitudinal distribution during the cycle, known as the
"butterfly diagram". Discovered by Hale (1919) the flip-flop
phenomenon of the global solar magnetic field with a period of 22
years showed the connection between variations of the sunspots
number and the magnetic field evolution. Taking into account the
Hale's law Bracewell (1953) proposed to include the information
about the periodical reversals of the magnetic field in the sunspot
number series by assigning alternating positive and negative signs to
the sunspot cycles, thus representing the sunspot cycles  as a
22-year oscillation. Bracewell suggested that the nonlinear
temporal distribution of the sunspots number can be written in the
form of a three-halfs law  \citep{brace88}: $W(t)=100\left[\mid
W_{lin}(t)\mid / 83\right]^{3/2}$, where $W_{lin}(t)$ is a new
"linearized" sunspot number, which is supposedly proportional to a
typical strength of the Sun's toroidal magnetic field.

For an explanation of the magnetic field generation Parker (1955)
proposed a simple dynamo model, which describes the phenomenon as an
action of two factors: the differential rotation and cyclonic
convective vortices. The mean-field theory and discovery of the
$\alpha$-effect give us a general description of the process of
magnetic field generation  \citep{krause80}. For recent reviews
we refer to Charbonneau (2005) and Brandenburg and Subramanian (2005).

It is known that the dynamo process is characterized by algebraic
and dynamic nonlinearities \citep[e.g.][]{klee07, sokol07}.
The algebraic nonlinearity can be determined as influence of the
magnetic field on fluid motions and on the kinetic helicity. This
results in quenching of the electromotive force and limits the growth
of the magnetic field. The evolution of the small-scale magnetic
helicity in the turbulent plasma causes a dynamical nonlinearity
in the dynamo process.

The turbulent helicity conditionally can be divided into two parts:
hydrodynamic and magnetic. The kinetic helicity describes helical
turbulent fluid motions; the magnetic helicity determines the order
of twisted magnetic field lines.  Due to the fact that the kinetic
helicity makes the magnetic field small-scaled, the back influence on
the turbulent fluid motions can restrict the unlimited growth of the
magnetic field. In the mean-field approach the magnetic helicity is
separated into large- and small-scale components. Because of
the conservation of the total helicity a growth of the large-scale
magnetic helicity due to the dynamo action is compensated by the
growth of the small-scale helicity of opposite sign \citep{sokol07}.
Thus, the small- and large-scale magnetic fields grow together
and are mirror-asymmetrical. This means that the condition of
magnetic helicity conservation is, perhaps, more severe for a
restriction of the dynamo action than the condition of the energy
conservation, which leads to quenching of the kinetic helicity.
Since the concept of magnetic helicity was introduce into the
dynamo theory by Pouquet {\it et al.} (1976) many magnetic helicity
models were suggested. For modeling the solar cycle we choose the
formulation of Kleeorin and Ruzmaikin (1982), explicitly based on
the idea of magnetic helicity conservation. Observational data of
solar magnetic fields are in a reasonable agreement with the idea of
the magnetic helicity conservation \citep{kleeorin2003, zhang2006,
sokol07, sokoloff2007}. Similar dynamo models with magnetic helicity
have been also considered in the context of galactic dynamos \citep{klee00}
and have included helicity transport by mean flows \citep{sur07}, which
            prevents the effect of `catastrophic' helicity
            quenching (e.g. Brandenburg and Subramanian, 2005).
            In our model, we don't include the helicity
            transport and consider the helicity quenching with
            a free parameter.

In this paper, we consider nonlinear behavior of the Parker's dynamo
model (without magnetic helicity) and the Kleeorin-Ruzmaikin model
(with magnetic helicity), and show that the latter can reproduce the
qualitative behavior of the sunspot number variations during the
solar cycles, including periodic and chaotic solutions for conditions
of the solar convective zone. For simplicity, we use a "low-order model"
approach \citep[e.g.][]{weiss84, soknef07}, reducing the dynamo equations
to a simple nonlinear dynamical system. In section 2, for consistency,
we reproduce the formulations of the Parker's and Kleeorin-Ruzmaikin's
models, and discuss conditions of linear stability and solar parameters.
In section~3, we present the numerical solutions and compare these with
observed properties of the sunspot number variations in solar cycles.

\section{Formulation}

\subsection{Parker's migratory dynamo}

In a kinematic approximation the dynamo problem can be described by
the induction equation \citep{parker55}
\begin{equation}
\frac{\partial\bf{B}}{\partial t} = \nabla \times ({\bf v \times B}) +
\eta_{m}\nabla^2 \bf{B},
\end{equation}
where $\bf{B}$ is the vector of magnetic field, ${\bf v}$ is the
vector of fluid velocity, $\eta_{m} =1/(\mu \sigma )$ is the
molecular magnetic diffusivity. The magnetic field, ${\bf B}$, and
the fluid velocity, ${\bf v}$, can be separated in two components
representing mean and fluctuating (turbulent) parts, or ${\bf B =
\left< B \right>+ b}$ and ${\bf v =\left< v \right>+ u}$. Here
$\left<{\bf B} \right>$ represents the averaged over longitude
magnetic field, ${\bf b}$ is the fluctuating part of ${\bf B}$,
$\left<{\bf v}\right>$ represents mean global-scale motions in the Sun,
(such as the differential rotation), ${\bf
u}$ is velocity of turbulent convective motions. Taking into account
that  the average of fluctuations is zero ($\left<{\bf b}\right> =
{\bf 0}$ and $\left<{\bf u}\right> = {\bf 0}$) for the case of isotropic
turbulence we obtain the following mean-field induction equation
\citep[e.g.][]{moff78}
   \begin{equation}
      \frac{\partial \left<{\bf B}\right>}{\partial t}=\nabla \times \left(\left<{\bf v}\right>\times\left<{\bf B}\right> +\alpha\left<{\bf B}\right> -\eta  \nabla \times \left<{\bf B}\right>\right),
   \end{equation}
where $\eta$ describes the total magnetic diffusion, which is the
sum of the turbulent and molecular magnetic diffusivity, $\eta
=\eta_{t}+\eta_{m}$ (usually $\eta_{m} \ll \eta_{t}$); parameter
$\alpha$ is helicity. The first term of the equation describes
transport of magnetic field lines with fluid, the second term
describes the $\alpha$-effect, and the last term determines diffusion
and dissipation of the field.

For describing the average magnetic field,  following
\citet{parker55}, we choose a local coordinate system, $xyz$, where axis
$z$ will represents the radial coordinate, axis $y$ is the azimuthal
coordinate and axis $x$ coincides with colatitude (figure 1). Effects
of sphericity are not included in this model. Hence, the vector of
the mean field, $\left<{\bf B}\right>$, can be represented as
\begin{equation}
\left<{\bf B}\right>=B_{y}(x,y){\bf e}_{y}+\nabla \times
\left[A(x,y){\bf e}_{y}\right],
\end{equation}
where $B_{y}(x, y)$ is the toroidal component of magnetic field,
$A(x, y)$ is the vector-potential of the poloidal field. Assuming
that $\left<{\bf v}\right> = \upsilon_{y}(x){\bf e}_{y} $ (rotational
component) we can write the dynamical system describing Parker's
model of the $\alpha$-dynamo \citep{parker55} in the standard form:
\begin{eqnarray}
\frac{\partial A}{\partial t}&=&\alpha B+\eta \nabla ^{2}A \\
\frac{\partial B}{\partial t}&=&G\frac{\partial A}{\partial x}+\eta
\nabla ^{2}B,
\end{eqnarray}
where $G = \partial\left<\upsilon_{y}\right>/\partial z$ is the rotational
shear. Here for simplicity we omit the subscript of $B$.

\begin{figure}[t]
\centerline{\includegraphics[width=10pc]{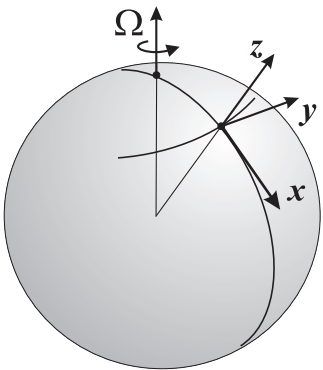}} \caption{Local
Cartesian reference coordinate system.}
\end{figure}

Assuming that the coefficients are constants and seeking a solution
of the model in the form $(A, B_{y})\sim(A_{0},B_{0})\exp[{\rm i}(kx-\omega t)]$,
we find the well-known result that a pure  periodic solution exists if
\begin{eqnarray}
D=\frac{\alpha G}{\eta^{2}k^{3}}= 2,
\end{eqnarray}
where $D$ is the so-called "dynamo number". The solutions grow in time for
$D > 2$, and decay for $D < 2$. As described in section 3, in the one-mode
approximation the classical Parker's dynamo model even in nonlinear cases
gives only non-chaotic oscillatory solutions, and, therefore, cannot explain
the observed variations of  the sunspot number in the solar cycles. For creating
chaotic variations of the magnetic field in the low-mode approximation it is
necessary to add  to the Parker's model a third equation describing variations
of the magnetic helicity and its interaction with the large-scale magnetic
field \citep{klee82, klee95}. We note that the absence of chaotic solutions
in the Parker's model with a simple algebraic quenching may be caused by the
low-order truncation \citep{schmalz91}, and that in a general PDE formulation
chaotic solutions may appear \citep{covas97, covas98}.

\subsection{The Kleeorin-Ruzmaikin model}

In this section for clarity we reproduce the derivation of the equation
for the magnetic helicity variations following \citet{klee82}.
For this we consider helicity $\alpha$ as a variable consisting of
two parts, hydrodynamic ($\alpha_{h}$)
and magnetic ($\alpha_{m}$) \citep{pouquet76, klee95}:
\begin{equation}
\alpha = \alpha_{h}+\alpha_{m},
\vspace*{-0.5cm}
\end{equation}
\begin{equation}
\alpha_{h}= -\tau\left<{\bf u}\cdot (\nabla \times {\bf u})\right>/3, \qquad \alpha_{m}= \tau\left<{\bf b}\cdot (\nabla \times {\bf b})\right>/(12\pi \rho),
\end{equation}
where $\tau \sim l_{0}^{2}/\eta _{t}$ is the lifetime of turbulent
eddies, $l_{0}$ is the characteristic scale of turbulent motions
in the convective zone, and $\eta_{t}$ is the turbulent diffusion
coefficient. It is convenient to define the influence of the
magnetic helicity on magnetic field using spectral density $\chi
(k)$ \citep{klee82}
\begin{equation}
\bar \chi \equiv \left<{\bf a \cdot b}\right>=\int \chi (k){\rm d}k,
\end{equation}
where ${\bf a}$ is the fluctuating part of the vector-potential of
magnetic field, ${\bf A}$, and $k$ is a wavevector.

To derive an equation for the averaged helicity density we
multiply the basic induction equation written without the
differential rotation term
\begin{equation}
\frac{\partial {\bf B}}{\partial t} = \nabla \times ({\bf v \times
B} - \eta_{m}\nabla \times {\bf B}),
\end{equation}
by the fluctuating part of the vector potential, $\bf{a}$; and also
multiply the equation for the vector-potential
\begin{equation}
\frac{\partial {\bf A}}{\partial t} = {\bf v \times B} -
\eta_{m}\nabla\times \nabla  \times {\bf A},
\end{equation}
by the fluctuating part of magnetic field ${\bf b}$, and finally sum
these equations. Then, using the averaging we obtain
\begin{eqnarray}
\frac{\partial \bar\chi}{\partial t} = \left<{\bf a} \cdot\frac{\partial{\bf
b}}{\partial t}+{\bf b} \cdot\frac{\partial{\bf a}}{\partial t}\right>=
\left<{\bf a  \cdot (\nabla \times (v\times \left<B\right>))}\right>+
\left<{\bf a  \cdot (\nabla \times(v\times b))}\right>+
\left<{\bf b  \cdot (v\times \left<B\right>)}\right>+\\ \nonumber
+\left<{\bf b  \cdot (v\times b)}\right>-
\eta _{m}{\bf \left(\left<a \cdot (\nabla \times (\nabla \times b))\right>+
\left<b  \cdot (\nabla \times b))\right>\right).}
\end{eqnarray}
Taking into account that ${\bf b=\nabla \times a}$, after some
transformations we obtain the following expression for the helicity
density
\begin{equation}
\frac{\partial \bar\chi}{\partial t}=-2 \left<[{\bf v\times b}]
\cdot\left<{\bf B}\right>\right>-2\eta_{m}\left<{\bf b} \cdot \nabla \times
{\bf b}\right>.
\end{equation}
Two terms $\left<\triangle[{\bf a \times [ v\times\left<B\right>}]]\right>$ and
$\left<\triangle[{\bf a\times [v\times b}]]\right>$, disappear as a result of
averaging over the volume.  Using the mean-field electrodynamics approximation
and retain only the first two terms for the mean electric field \citep{moff78}
\begin{equation}
\bf{\varepsilon\equiv \left<v\times b\right>} \cong
\alpha\left<\bf{B}\right>-\eta\left(\nabla \times \left<\bf{B}\right>\right),
\end{equation}
we obtain
\begin{equation}
\frac{\partial \bar\chi}{\partial
t}=2\left(\eta\left<\bf{B}\right> \cdot (\nabla \times\left<\bf{B}\right>)-
\alpha\left<\bf{B}\right>^{2}-\eta_{m}\left<\bf{b} \cdot\nabla \times
\bf{b}\right>\right).
\end{equation}

Then we find a relationship between $\alpha_{m}$ and $\bar\chi$. For
this we assume that in inertial range $(k_{0}, k_{1})$ the last
term of (15) can be written as
\begin{equation}
\left<\bf{b} \cdot (\nabla \times \bf{b})\right>=\int_{k_{0}}^{k_{1}}k^{2}\chi(k){\rm d}k,
\end{equation}
where the helicity  density spectrum, $\chi(k)$, can be described as
\citep{klee82, klee95, zeld83}:
\begin{equation}
\chi(k)=\bar\chi\frac{q-1}{k_{0}}\left(\frac{k_{0}}{k}\right)^{q}\left[1-
\left(\frac{k_{0}}{k_{1}}\right)^{q-1}\right]^{-1}.
\end{equation}
Here $k_{0}^{-1}$ is an outer scale of a turbulence, $k_{1}^{-1}$ is
the scale of the helicity spectrum cutoff. The value of parameter
$q$ depends on the type of nonlinear interactions in the turbulent
plasma. For example, $q=5/3$ describes the case of the Kolmogorov
spectrum, and $q=3/2$ corresponds to turbulence of interacting Alfv\'en
waves (Kraichnan's spectrum). In addition, factor $q$ was selected
in such a way that coefficient $\chi$ coincides with the mean helicity.
Consequently, the magnetic helicity given by (8)
becomes
\begin{equation}
\alpha_{m}=\frac{\tau}{12\pi \rho}\left<\bf{b} \cdot (\nabla \times
\bf{b})\right>=\frac{1}{12\pi \rho}\int_{k_{0}}^{k_{1}}k^{2}\tau
_{*}(k)\chi(k){\rm d}k,
\end{equation}
where $\tau_{*}=2\tau_{0}(k/k_{0})^{1-q}$, $\tau_{0}$ is relaxation
time, averaged over the energy spectrum.

Taking into account that $|\bar\chi|=|\left<\bf{ab}\right>|\sim
B^{2}/k_{0}$ and $\eta_{t}=\left(12\tau_{0}k_{0}^{2}\right)^{-1}$,
and substituting (17) into (18) we obtain
\begin{equation}
\alpha_{m}=I\bar\chi,
\end{equation}
where
\begin{equation}
I=\frac{1}{18}\left(\frac{q-1}{2-q}\right)\frac{1}{4\pi\rho\eta_{t}}
\left(\frac{k_{*}^{4-2q}-1}{1-1/k_{*}^{q-1}}\right)
\end{equation}
and  $k_{*}=k_{1}/k_{0}$.

Finally, multiplying equations (15) and (20) we obtain the expression for
variations of the magnetic helicity in terms of the mean magnetic
field:
\begin{equation}
\frac{\partial \alpha_{m}}{\partial t}=\frac{Q}{2\pi \rho}\left[\left<{\bf B}\right>  \cdot
(\nabla \times \left<{\bf B}\right>)-\frac{\alpha}{\eta}\left<{\bf B}\right>^{2}\right]-\frac{\alpha_{m}}{T},
\end{equation}
where
\begin{eqnarray}
Q=\frac{1}{18}\left(\frac{q-1}{2-q}\right)\left(\frac{k_{*}^{4-2q}-1}{1-1/k_{*}^{q-1}}\right), \qquad
T=\frac{1}{2\eta_{m}k_{0}^{2}}\left(\frac{3-q}{q-1}\right)\left(\frac{1-1/k_{*}^{q-1}}{k_{*}^{3-q}-1}\right).
\end{eqnarray}
Equation (21) is written for the case of an uniform turbulent
diffusion ($\eta_{t}=(6\tau_{0}k_{0}^{2})^{-1}$), and when the magnetic
Reynolds number is large, $\eta \approx \eta_{t}$. In this model
we do not include transport of magnetic field by flows.

For further analysis of the Kleeorin-Ruzmaikin model we transform
equations (4)-(5) and (21) in a nonlinear dynamical system in
non-dimensional variables. Following the approach of Weiss {\it et
al.} (1984) we average the system of equations (4 - 5 and 21) in a
vertical layer to eliminate $z$-dependence of $A$ and $B$ and
consider a single Fourier mode propagating in the $x$-direction
assuming $A=A(t)e^{{\rm i}kx}, B=B(t)e^{{\rm i}kx}$; then we get the following
system of equations
\begin{eqnarray}
\frac{{\rm d} A}{{\rm d} t}&=&\alpha B-\eta k^{2} A, \qquad \frac{{\rm d} B}{{\rm d} t}={\rm i}kG A-\eta k^{2}B,\nonumber \\
\frac{{\rm d} \alpha_{m}}{{\rm d} t}&=&-\frac{\alpha_{m}}{T}-\frac{Q}{2\pi \rho}
\left[-ABk^{2}+\frac{\alpha}{\eta}\left(B^{2}-k^{2}A^{2}\right)\right].
\end{eqnarray}

This transformation allows us to investigate more easily various
nonlinear regimes, from periodic to chaotic, and obtain
relationships of the basic properties, such as the cycle growth and
decay times, duration and amplitude. However, we note that the
formulation and the interpretation of solutions of the simplified
system are not straightforward because it does not adequately
describes nonlinear coupling of the spatial harmonics. Recently,
\citet{soknef07} developed a self-consistent method of reducing the
dynamo equations, (10) and (11), to a dynamical system. In
particular, it follows from their study that the dominant modes of
the toroidal field in the case of the solar dynamo are described by
the harmonics, which are antisymmetric with respect to the equator
(in accordance with the Hale law), $\sin(kx)$, where $x$ is
colatitude, and the wavenumber, $k$, is even: $k=2, 4, ...$ The
first $k=2$ harmonic has the largest growth rate.
We retain only this mode in our dynamical model. This
nonlinear dynamical system is solved numerically by using standard
{\it Mathematica} software for high-precision integration of
potentially stiff problems.

For interpretation of solutions of the dynamical system in
terms of the sunspot number properties we use the imaginary part of
the toroidal component $B(t)$ because it gives the amplitude of
the antisymmetric harmonics, and approximate the sunspot number, $W$, as
$({\rm Im}B)^{3/2}$, following Bracewell's suggestion. Of course,
there might be different definitions of the relationship between
characteristics of the magnetic field and the sunspot number
parameter. We note that the solutions of the dynamical system are
qualitatively similar for the different harmonics. Nevertheless, we
choose the parameters, which correspond to the solar situation.

Making the following substitutions: $A=A_{0}\hat A$, $B=B_{0}\hat
B$, $t=T_{0} \hat t$, $k=\hat k/r$ ($r$ is a layer radius),
$T_{0}=1/(k^{2}\eta)$ and $\alpha_{m}=\alpha_{0}\hat \alpha_{m}$,
and taking into account that $A_{0}=B_{0} \eta k /G$ we
obtain:
\begin{eqnarray}
\frac{{\rm d} \hat A}{{\rm d}\hat t}&=&\hat D\hat B-\hat A,
\qquad
\frac{{\rm d}\hat B}{{\rm d}\hat t}={\rm i}\hat A-\hat B, \nonumber \\
\frac{{\rm d}\hat \alpha_{m}}{{\rm d}\hat t}&=&-\nu \hat\alpha_{m}
+ \left[\hat A \hat B - \hat D \left(\hat B^{2}-\lambda \hat
A^{2}\right)\right],
\end{eqnarray}
where $\hat D = D_{0}\hat \alpha$ and $\hat \alpha =\hat
\alpha_{h}+\hat \alpha_{m}$ are the non-dimensional dynamo number and
helicity, $D_{0}=\alpha_{0}Gr^{3}/\eta^{2}$,
$\alpha_{0}=2Qk\upsilon_{A}^{2}/G$, $\upsilon_{A}$ is the Alfv\'en
speed, $\nu = T_{0}/T$ (here we assume that $k_{0}$ is close to $k_{1}$
then $T\sim 1/(2k_{0}^{2}\eta_{m}), $ \citep{klee82}) and
$\lambda=(k^{2}\eta/G)^{2}={\rm Rm}^{-2}$, $k$ is a characteristic
wavelength, Rm is the magnetic Reynolds number. In the next section we
discuss the ratio of parameters when the variations become
autonomous.

\subsection{Linear stability}

First, we simplify the third equation of system (24) by neglecting
the last term, because $\lambda$ is very small for large magnetic
Reynolds numbers. Note, that hereafter we omit the hat symbol for
all non-dimensional variables. Hence (24) becomes
\begin{eqnarray}
\frac{{\rm d} A}{{\rm d} t}&=& D B- A,  \qquad
\frac{{\rm d} B}{{\rm d} t}={\rm i} A - B, \nonumber \\
\frac{{\rm d} \alpha_{m}}{{\rm d} t}&=&-\nu \alpha_{m} +\left[AB -
D B^{2}\right].
\end{eqnarray}
Using a linearization procedure, $A=A'+A^{0}$,  $B=B'+B^{0}$ and
$\alpha=\alpha_{m}'+\alpha_{h}$, and taking into the account that
the dynamo number is $D=D_{0}\alpha$, we obtain the following
equations
\begin{eqnarray}
\frac{{\rm d} A'}{{\rm d} t}&=& D_{0}\alpha_{h}B^{0}- A',
\qquad \frac{{\rm d} B'}{{\rm d} t}={\rm i} A' - B', \nonumber \\
\frac{{\rm d} \alpha_{m}'}{{\rm d} t}&=&-\nu \alpha_{m}'
+\left[A^{0}B'+A'B^{0}-2D_{0}\alpha_{h}B'B^{0}-2D_{0}\alpha_{m}'B_{0}^{2}\right].
\end{eqnarray}
Then, considering solutions in the form $\sim e^{-{\rm i}\omega t}$ we
obtain the condition of linear stability: $|D_{0}\alpha_{h}|> 2$, in
the approximation of $B^{0} = 0$ and $A^{0}=0$. Thus, in the case of
$|D_{0}\alpha_{h}|< 2$ the magnetic field does not grow. For
$|D_{0}\alpha_{h}|=2$ we have a periodical solution like in the
Parker's model. The linear theory also provides the direction of the
migration of the dynamo waves with latitude. These waves travel from
higher to lower latitudes if $D$ is negative, and in the opposite
direction for positive $D$. Thus, to make the model consistent with
the butterfly diagram we consider the case of negative $D$.

\subsection{Solar parameters}

In order to estimate the range of parameters of the
Kleeorin-Ruzmaikin model, and for modeling the solar
cycle we used the standard model of the interior structure of the
Sun for the top, bottom and middle areas of the convective zone
(Table 1).
\begin{table}[b]
  \tbl{Parameters of the standard model for the different parts of the solar convective zone.}
{\begin{tabular}{@{}lccc}\toprule
   Parameter  & bottom & middle & top$^{\rm *}$ \\
\colrule
radius ($r$), cm &  $5 \times 10^{10}$&  $6 \times10^{10}$ &$7 \times10^{10}$ \\
density ($\rho$), g/cm$^{3}$ & $2 \times10^{-1}$ & $4 \times10^{-2}$  & $2 \times10^{-3}$  \\
turbulent diffusivity ($\eta_{t}$), cm$^{2}$/s & $2 \times10^{13}$ & $3 \times 10^{13}$  & $1 \times 10^{13}$  \\
radial velocity gradient ($G$)$^{\rm **}$, s$^{-1}$ & $5\times 10^{-6}$ & $2\times 10^{-6}$ & $-2\times 10^{-6}$\\
   \botrule
  \end{tabular}}
 \tabnote{$^{\rm *}$ Supergranulation layer.}
  \tabnote{$^{\rm **}$ Calculated from the helioseismology inversion results of \citet{Schou1998}}
\end{table}
The key parameter of the model is the dynamo number $D=D_{0}\alpha$,
because its magnitude determines behavior of the magnetic field.
We remind that according to the condition of linear stability,
$D_{0}\alpha_{h}$ should be greater than 2. Taking into account that
$D_{0}=\alpha_{0}Gr^{3}/\eta^{2}$, we assume that $\eta \approx
\eta _{t}$, $G\sim \left<\upsilon_{y}\right>/r$, where $\left<\upsilon_{y}\right>$
is a typical rotational velocity, $r$ is the radius of convective layers.
Parameter $\lambda$ determines the influence of vector-potential $A$ on
variations of  magnetic helicity $\alpha_{m}$. From our estimates it follows
that for the solar conditions $\lambda \leq 10^{-4}$. Consequently, we can
neglect the term with $\lambda$. The last non-dimensional parameter, $\nu$,
included in equations (24) describes the ratio of two characteristic times, $T_{0}$
and $T$. Parameter $T_{0}=1/(k^{2}\eta)$ is estimated using the values of
turbulent diffusivity $\eta_{t}=1.5H_{p}V_{conv}$ from the mixing-length model,
where $H_{p}$ is the pressure scale height, $V_{conv}$ is the convective velocity.
In the absence of helicity fluxes the value of the damping parameter $\nu$
   is small. However, in reality the helicity fluxes increase the dissipation rate.
   This can be modeled as a damping term, which increases
   the effective value of $\nu$ \citep{blackman03}.
   Because of this the value of $\nu$ is to some extent uncertain.

\section{Numerical calculations}

\subsection{Periodic solutions}

According to the model of Parker (equations 4 and 5), the pure harmonic
variations of the magnetic field occur for the dynamo-number $D =
2$. It is also known, that for $|D| >2$ we have solutions with
increasing amplitude. Following a standard procedure, we included a
nonlinearity (alpha-quenching) as $\alpha_{h}/(1+\xi B^{2})$ \citep{ivanova77,brand05},
where  $\xi$ is a quenching
parameter, which limits the growth of the magnetic field amplitude.
Examples of nonlinear solutions  for the Parker's model are shown in figure 2.
However, our numerical calculation showed that in the nonlinear regime
the variations of the
toroidal field (figure 2a) and the sunspot number (figure 2b) are
also periodic and similar to the classical case of the linear harmonic
solution for $|D|=2$. For calculations of the model sunspot number
we used Bracewell's suggestion (private communication) of the
connection between the toroidal field strength, $B$, and the sunspot
number, $W$, in the form of the three-halfs law: $W \sim B^{3/2}$.

We note that the model shows a larger amplitude of variations for
the poloidal component than for the toroidal field, and that there is
a phase shift between them. However, comparing of the observed
profiles of the solar cycles (figure 9) with the solutions for the
Parker model (figure 2b), we see that, in general, the model
solutions have a sinusoidal character and do not describe the basic
properties of the sunspot profiles with rapid growth and slow decay.

\begin{figure}[t]
\centerline{\includegraphics[width=27pc]{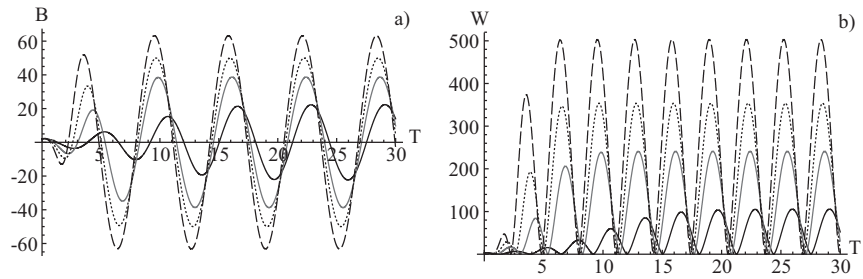}}
\caption{Solutions (in non-dimensional units) for the toroidal magnetic field (a) and the
sunspot number (b) in case the Parker's dynamo model for different
values of the dynamo number, $D$, and quenching parameter $\xi =
10^{-3}$: $D = 3$ (solid black curve), $D = 5$ (grey curve), $D = 7$
(dotted curve) and $D = 10$ (dashed curve).}
\end{figure}

For the Kleeorin-Ruzmaikin model, given by equations (24), the linear
instability condition is also $|D|\equiv|\alpha_{h}D_{0}|> 2$.
However, in this case the profile of the periodic solutions is not
sinusoidal, and depends on the initial conditions, $A_0$ and $B_0$.
For higher initial values the amplitude of the nonlinear
oscillations in the stationary state is higher. However, the shapes
of the oscillation profiles are similar.

Figure 3 illustrates solutions for the model of
Kleeorin-Ruzmaikin, and the corresponding variations of the sunspot number
for different initial conditions. As mentioned, changes of initial values
for magnetic field components $A_{0}$ and $B_{0}$ leads to very similar
profiles. In high amplitude cases, dual peaks may appear in the variations
of the vector potential, $A$, of the poloidal field. The evolution
of the magnetic helicity represents a relatively smooth growth followed
by a sharp decay. The helicity has maxima when the toroidal field is zero.
In these calculations the value of parameter $\nu$, which describes damping
rate of magnetic helicity and depends on the turbulence spectrum and the
dissipation though helicity fluxes, is of the order of unity. Finally,
the variations of the sunspot number, $W$, with the amplitude increase are
characterized by higher peaks and shorter rising times (figure 3d). Note that
in the sunspot number profile we can recognize the well-known general properties
of the sunspot number profile with the rapid growth at the beginning of a cycle
and a slow decrease after the maximum.

\begin{figure}[b]
\centerline{\includegraphics[width=35pc]{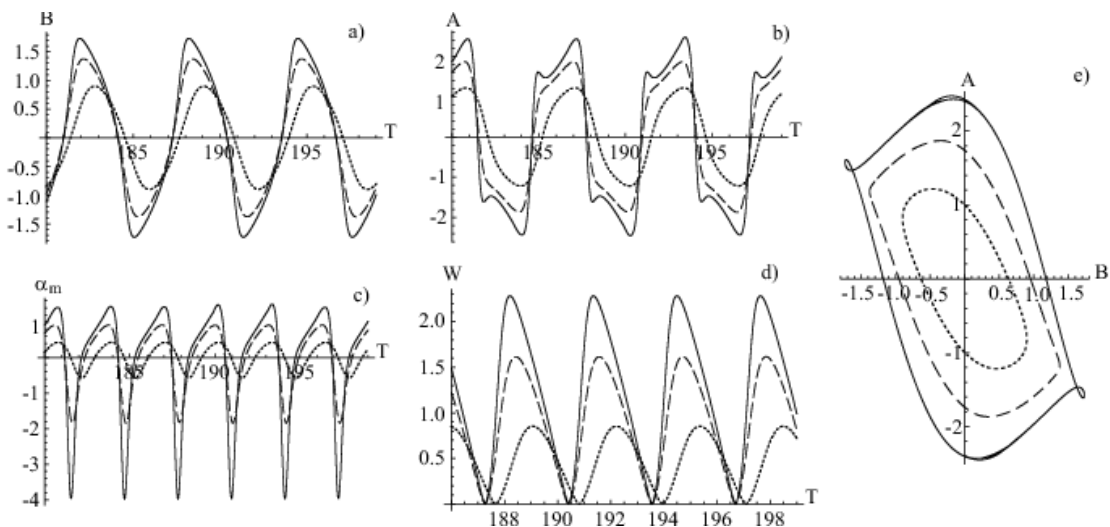}}
\caption{Variations of the magnetic field for the middle convective zone
$\alpha_{h}D_{0} = -2$: $\nu = 1.28$, $\alpha_{h}=2.439$, $D_{0}=-0.82$
for different initial conditions: $B_{0}=4{\rm i}$, $A_{0}=-0.01{\rm i}$ (dotted curve),
$B_{0}=4{\rm i}$, $A_{0}=-{\rm i}$ (dashed curve) and $B_{0}=1+4{\rm i}$, $A_{0}=-{\rm i}$
(black curve): a) toroidal component, $B$; b) vector-potential, $A$,
of the poloidal magnetic field; c) magnetic helicity variations;
d) evolution of the model sunspot number; e) phase portrait of the magnetic
field components.}
\end{figure}

With the increase of $|\alpha_{h}D_{0}|$  ($|\alpha_{h}D_{0}|>2$)
the profile of magnetic field variations continue to deform and can
become unstable with very steep variations of the magnetic field.
The solution can be stable again if we enhance the back reaction by
increasing the quenching parameter. We use the following quenching
formula for the kinetic part of helicity, $\alpha_h$ \citep{klee95}
\begin{equation}
\alpha =\frac{\alpha_{h}}{1+\xi B^2}+\alpha_{m}.
\end{equation}
Thus we always have a possibility for selecting $\xi$ to obtain
periodic nonlinear solutions.
\begin{figure}[t]
\centerline{\includegraphics[width=30pc]{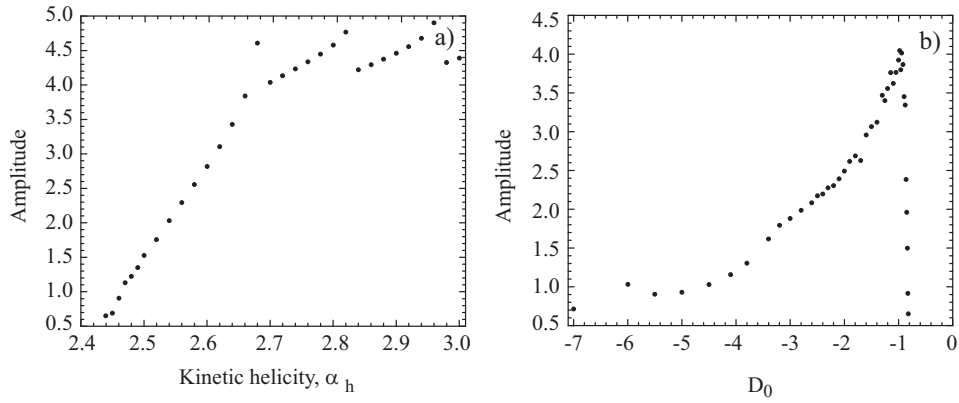}}
\caption{Dependence of the amplitude of model sunspot number $W$ on
a) kinetic helicity $\alpha_h$ at fixed dynamo number $D_0=-0.82$;
b) dynamo number $D_0$ at fixed $\alpha_h=2.44$.}
\end{figure}

\begin{figure}[b]
\centerline{\includegraphics[width=40pc]{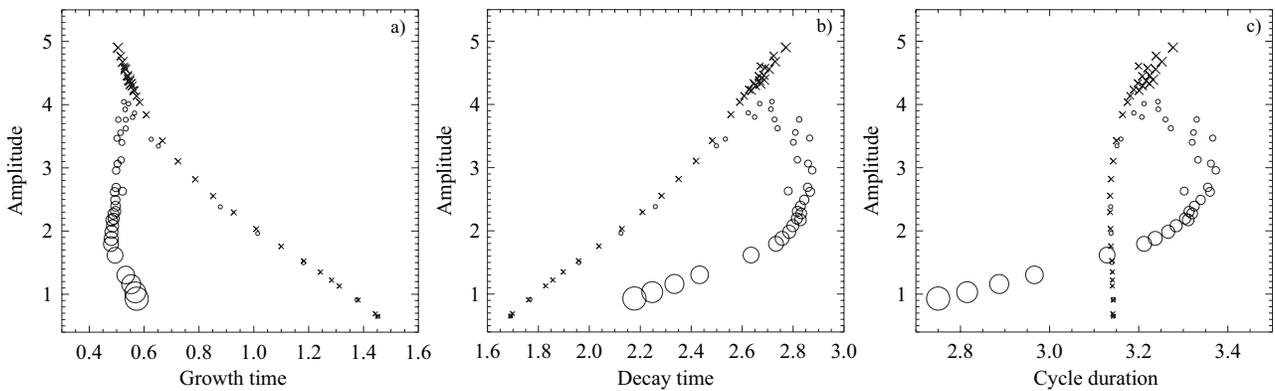}}
\caption{Relationships between the model sunspot number amplitude,
$W$ and a) the cycle growth time, b) the decay time, and c) the
cycle duration. The circles show a sequence for a fixed value of the
kinetic helicity, $\alpha_h=2.44$ and the dynamo number varying from
-7 to -0.82. The crosses show the case of fixed $D_{0}=-0.82$ and varying
$\alpha_{h}$, from 2.44 to 3. The size of the crosses and circles
is proportional to the corresponding values of $D_0$ and $\alpha_h$.
Other model parameters correspond to the middle convective zone:
$\nu=1.28, \lambda =1.23\times 10^{-6}$; and the initial conditions
are $A_{0}=-{\rm i}$, $B_{0}={\rm i}$, $\alpha_{m}^{0}=0$.}
\end{figure}

As mentioned earlier, the change in values of the dynamo number and
the kinetic helicity leads to a change in the amplitude of the
magnetic field variations (figure 4). With the increase of the kinetic
helicity, $\alpha_h$, above the linear stability level at a fixed
dynamo number (in this example, $D_0=-0.82$) the amplitude rapidly
grows (figure 4a). However, for a fixed helicity value (e.g. $\alpha_h=2.44$)
and increasing value of dynamo number $D_0$ the amplitude rapidly grows
and then sharply declines (figure 4b). For other values of $D_0$ and
$\alpha_h$, satisfying the linear instability condition, the solutions
are qualitatively similar.

Figure 5 shows the relationships between of the amplitude of the
sunspot number parameter, the growth and decay times and the cycle
duration, for different values of $\alpha_{h}$ and $D_{0}$. These
characteristic times were determined from the points of minima and maxima of the
model sunspot number, $W$. The crosses represent the amplitude of the
periodic field variations for $D_{0}=-0.82$  and different values of
kinetic helicity, $\alpha_{h}$. The circles correspond to the case
of constant $\alpha_{h}=2.44$ and different values of dynamo number
$D_{0}$. In the first case, the relationship between the cycle
amplitude and the growth time is well-defined and monotonic.
However, in the case of a fixed $\alpha_h$ and varying $D_0$ (circles)
the amplitude initially, at small $D_0$, follows the same sequence
as in the variable $\alpha_h$ case, but then at higher values of
$|D_0|$ (shown by bigger circles) the amplitude decreases. The decay
time (figure 5b) is longer for higher amplitude cycles. The
relationship between the amplitude and the cycle duration (figure 5c)
is more complex. In the case of  fixed $D_0$ the amplitude and
the cycle duration increase with the dynamo number, but initially at
lower $\alpha_h$, just above the stability value, the amplitude
rapidly grows but the duration does not change much. With
further increase of $\alpha_h$ the amplitude grows slowly, but the
cycles become longer. In the case of the varying dynamo number
(circles) the relationship is multivalued, that is the cycles of the
same duration may have different amplitudes.

\subsection{Chaotic solutions}
\begin{figure}[b]
\centerline{\includegraphics[width=27pc]{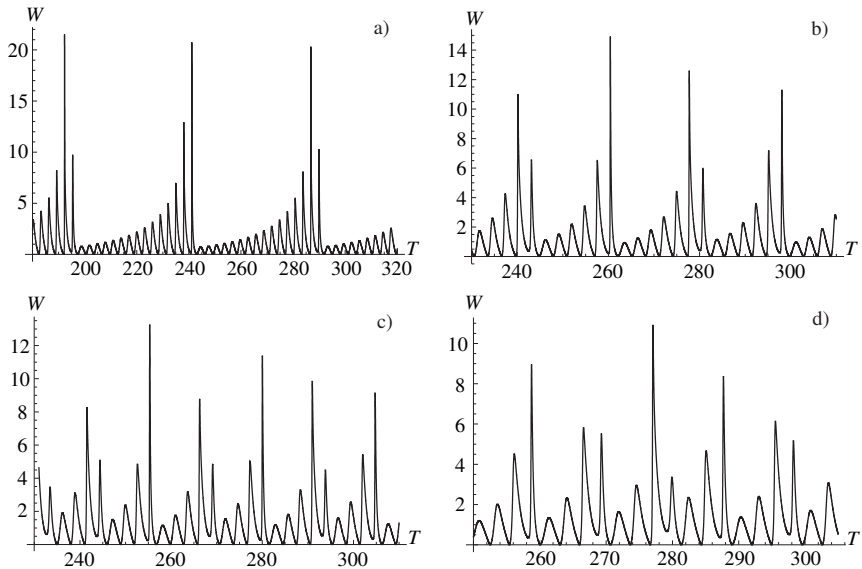}}
\caption{Variations of the model sunspot number, $W$, illustrating
the transition from the periodic to chaotic solutions for
conditions of the middle convective zone, dynamo number $D_0=-0.82$,
$\nu=1.28$, $\lambda=1.23\times 10^{-6}$ and different values of the
kinetic helicity $\alpha_{h}$ ($D_{0}\alpha_{h}$): a) $\alpha_{h}=2.58$,
$\xi=10^{-3}$ (-2.116), b) $\alpha_{h}=2.8$, $\xi=10^{-3}$ (-2.296),
c) $\alpha_{h}=3$, $\xi=1.3\times10^{-3}$ (-2.46),  and
d) $\alpha_{h}=3.2$, $\xi=3.9\times10^{-3}$ (-2.624).}
\end{figure}

The transition of the periodic to chaotic solutions occurs when
the dynamo number, $|\alpha_{h}D_{0}|$, increases.
In the transition regime the cycle
amplitude becomes modulated: it slowly increases with time, and then suddenly and very
sharply declines, and then start growing again (figure 6a). Some
examples of the transition to the chaotic regime for the sunspot
number parameter and the development of chaotic behavior are shown
in figure 6 for a fixed $D_{0}=-0.82$ and increasing values of the kinetic
helicity, $\alpha_h$.

\begin{figure}[t]
\centerline{\includegraphics[width=37pc]{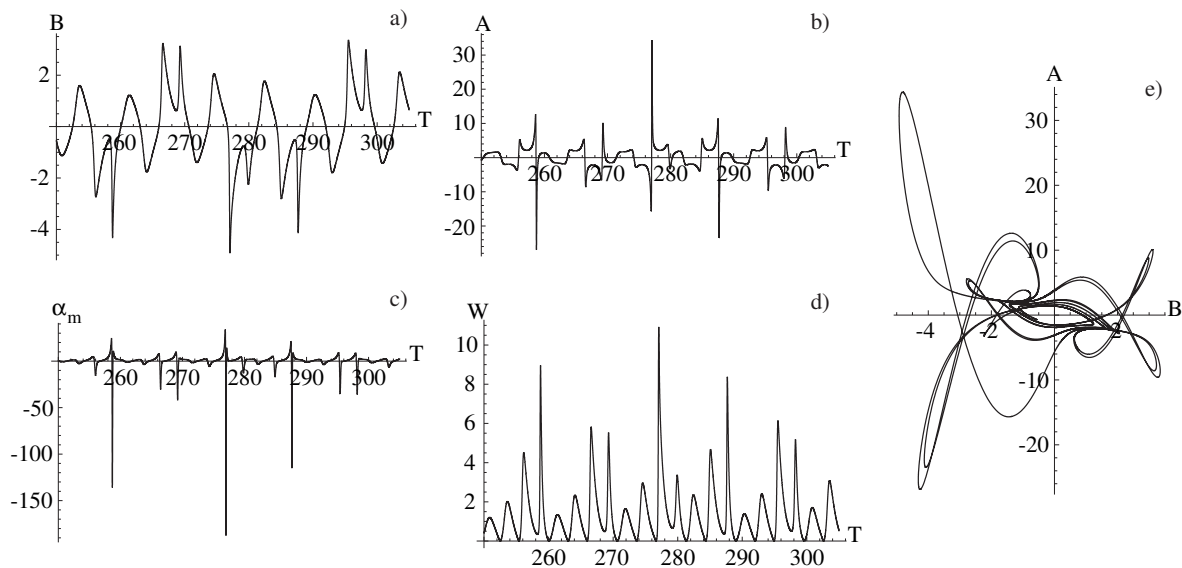}}
\caption{Example of the chaotic solution for the middle convective zone
($\nu=1.28$, $\lambda=1.23\times 10^{-6}$, $D_{0}=-0.82$,
$\alpha_{h}=3.2$, $\xi=3.9\times10^{-3}$): a) toroidal field $B$, b)
vector-potential $A$, c) magnetic helicity $\alpha_m$, d) model
sunspot number $W$ and e) the phase portrait of the magnetic
components.}
\end{figure}

At small deviation from the periodic regime the slow rise of
amplitude is followed by sharp recession. Nevertheless, each
subsequent increase in the amplitude is almost identical to the
previous one, or in other words, we have in this case, a modulated
periodic solution  (figures 6 a, b). Also, we note that the shape of
the profile for the each short cycle is still the same as in the
case of the periodic solutions with  the rapid growth and slow
decline. With further increase of the kinetic helicity, the general
behavior with growing cycle amplitudes and subsequent recessions
continues, but the variations of amplitude become increasingly more
chaotic (figures 6 c, d). The amplitude of the solutions is controlled
by the quenching parameter, $\xi$, which was chosen in figure 6 to obtain the
solutions with amplitudes of the same order of magnitude.

\begin{figure}[b]
\centerline{\includegraphics[width=40pc]{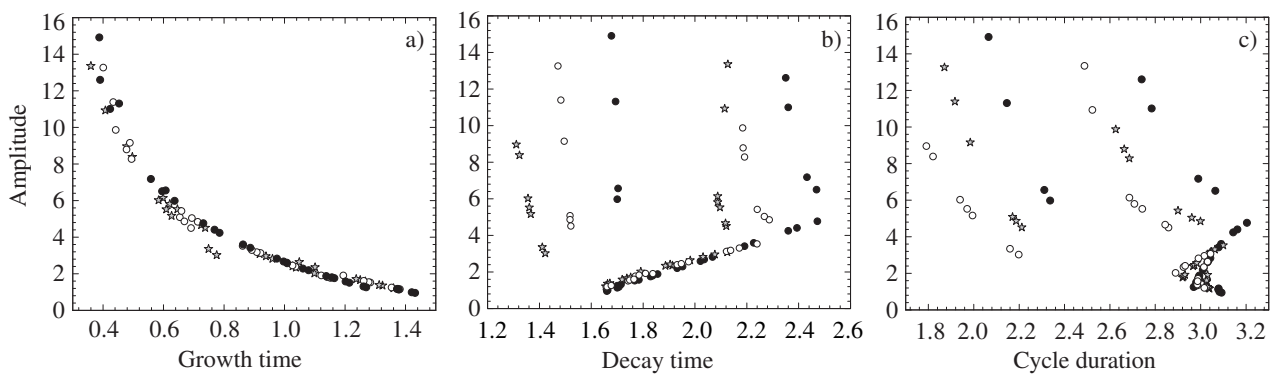}}
\caption{Relationships between the amplitude of the model sunspot
number and a) the growth time, b) decay time and c) the cycle
duration, for $D_{0}=-0.82$ and different values of the kinetic
helicity: $\alpha_{h}=2.8$ (black circles), $\alpha_{h}=3$ (empty
circles), $\alpha_{h}=3.2$ (stars). The time scales are non-dimensional
(in units $T_{0}=1/(k^2\eta)$).}
\end{figure}

In the case of significant deviations from the condition of linear
stability the solutions become chaotic for all variables of the
dynamical system. Figure 7 shows an example of chaotic variations
for the magnetic field components, the magnetic helicity and the
sunspot number parameter. In the chaotic solutions, the peaks of the
toroidal magnetic field, $B$ (figure 7a) strongly correlate with the
peaks of the vector-potential, $A$, and the magnetic helicity,
$\alpha_{m}$, (figures 7 b, c). The growth of the toroidal field also
leads to strengthening of the poloidal field and strong fluctuations
of the magnetic helicity. The comparison of the variations for the
toroidal component and vector-potential on the phase portrait (figure
7e) shows behavior resembling a strange attractor, when the
trajectory converges to a steady state, then deviates from it, and
returns again.

\begin{figure}[t]
\centerline{\includegraphics[width=30pc]{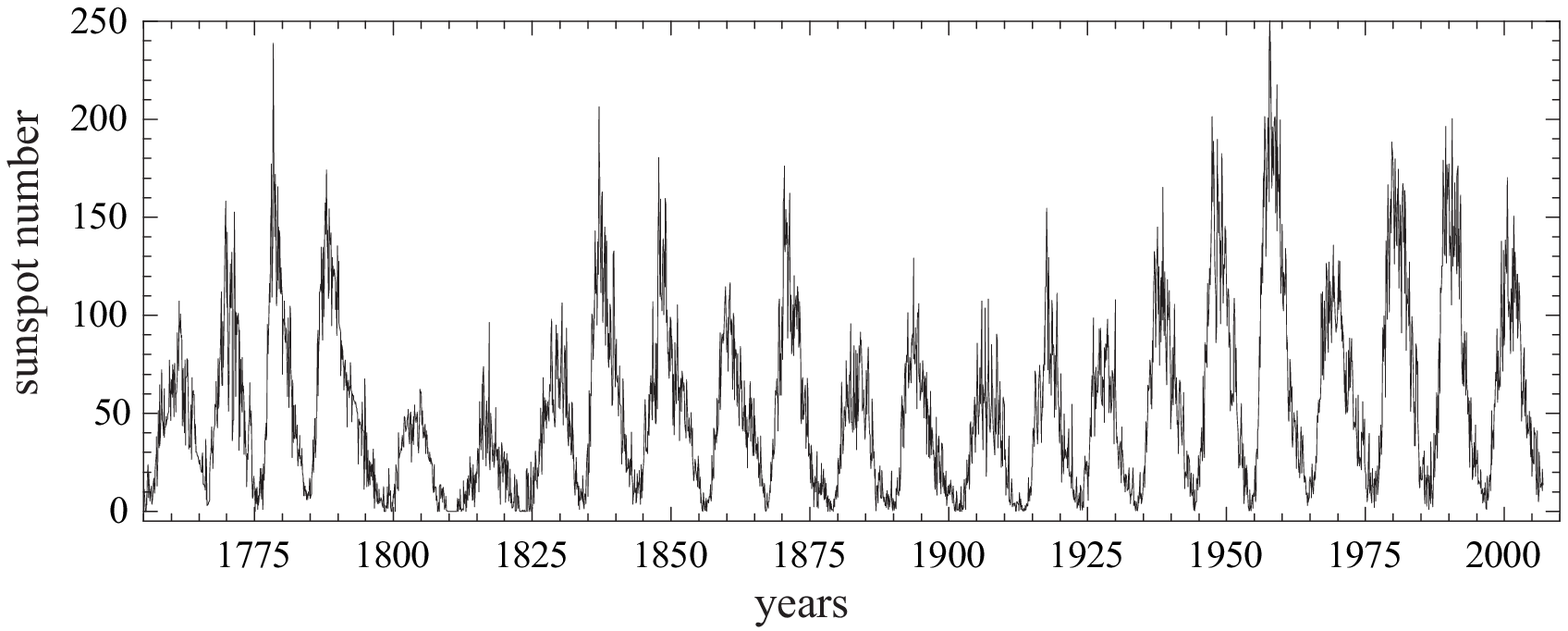}}
\caption{Observed monthly sunspot number series for 1755 - 2007 yrs. from
National Geophysical Data Center. }
\end{figure}

\begin{figure}[t]
\centerline{\includegraphics[width=35pc]{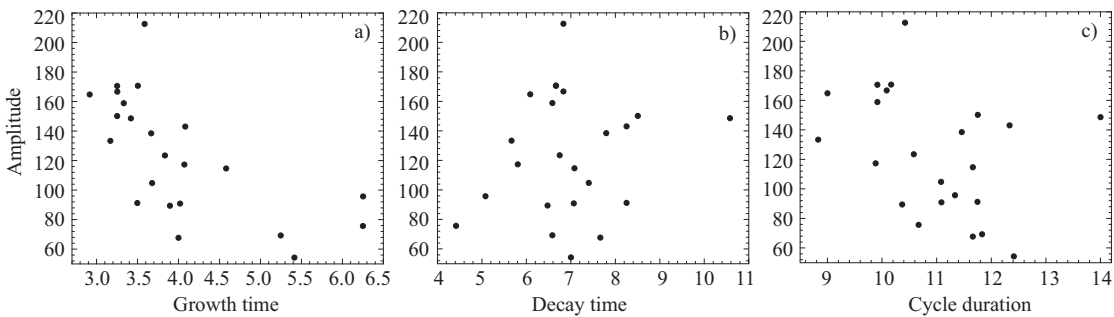}}
\caption{Properties of the solar cycles of 1755 - 2007: the
relationships between the amplitude of the sunspot number and a) the
cycle growth time, b) the cycle decay time and c) the cycle
duration. The time scales are in years.}
\end{figure}

The peaks ("cycles") of the chaotic solutions still reproduce the
typical profiles of the sunspot number variations described for the
periodic solutions. However, the duration and amplitude of the
cycles vary. For comparison in the figure 8 we show the properties
of the different cycles of three chaotic solutions for $D_{0}=-0.82$
and: $\alpha_{h}=2.8$ (black circles), $\alpha_{h}=3$ (empty circles)
and $\alpha_{h}=3.2$ (stars).
In particular, we can see that the growth time is shorter
for stronger cycle (figure 8a). The decay time, in general, is also
longer for stronger cycle (figure 8b). However, the relationship
between the cycle duration and its amplitude is again not very
certain (figure 8c).

It is interesting to note that the similar relationships between the
amplitude and the characteristic times, which we found for both
periodic and chaotic solutions of the Kleeorin-Ruzmaikin model,
occur for the real solar cycles (figures 9 and 10). For comparing the
properties of the simulated variations of the sunspot number and the
solar cycles we used the monthly data of the sunspot number (figure 9)
for the period of 1750 - 2007 from National Geophysical Data Center ({\it ftp://ftp.ngdc.noaa.gov/stp/solar\_data/sunspot\_numbers/monthly.plt}).
The growth and decay times were measured from the sunspot minimum to
the maximum and from the maximum to the next minimum, using the
same procedure as for the model. The results shown in figure 10 resemble the
theoretical relationships of figures 5 and 8. In particular, the growth
time is shorter for stronger cycles. The decay time and the duration
do not show a clear correlation with the amplitude.

\section{Discussion}

We have presented a numerical analysis of simple dynamical
models describing the nonlinear behavior of two dynamo models, the
classical Parker's dynamo model with the standard $\alpha$-quenching
and the model of Kleeorin and Ruzmaikin (1982), which takes into
account effects of turbulent magnetic helicity. The models are analyzed
in the low-mode approximation with the goal of representing variations of
a sunspot number parameter during the solar cycles. We found that the
Parker's model does not  reproduce the typical behavior of the sunspot
number with a fast growth and slow decay or obtain a chaotic solution
in the low-mode approximation, even in strong nonlinear regimes. The
analysis of the Kleeorin-Ruzmaikin model showed the existence of
nonlinear periodic and chaotic solutions for conditions of the solar
convective zone. For this model we obtained the profiles of the sunspot
number variations, which qualitatively reproduce the typical profile
of the solar cycles.

It is interesting that the properties of the simulated cycles
demonstrate good qualitative agreements with the properties of the
observed solar cycles. In particular, the relationship between the
cycle amplitude and the growth time looks similar. Of course, such
simplified models cannot pretend to provide a realistic description
of the solar cycles. More complicated models such as the models with
flux transport \citep[e.g.][]{dikpati06} should be considered for
more detailed description of the solar magnetic field evolution.

Nevertheless, these results encourage further theoretical and
observational investigations of the dynamo model with the turbulent
magnetic helicity. We also hope that through the use of a simple
dynamical model reproducing the basic properties of the solar cycle
it will become possible to apply the data assimilation methods for
predicting the solar cycles \citep[e.g.][]{kiti08}.


\label{lastpage}
\end{document}